\documentclass[prb,aps,twocolumn,showpacs]{revtex4}

\usepackage{graphicx}
\graphicspath{{Figures/},{Figures/src/}}      
\DeclareGraphicsExtensions{.eps,.ps}
\usepackage{amsmath}

\newcommand\Ef{E_F}

\begin{document}

\title{Discrete thinning dynamics in a continuum model of metallic nanowires}

\author{J.\ B\"urki}

\affiliation{Department of Physics, University of Arizona, Tucson, AZ 85721}

\date{December 13, 2006 (submission to PRB); accepted for publication March 27, 2007}

\begin{abstract}
Transmission electron microscopy experiments have recently observed gold metal nanocylinders to thin down in a discrete manner:
A kink---a step of order of one atomic layer---nucleates at one end and then moves along the wire, leaving a thinner cylinder behind it. 
In this paper, I show that a similar thinning process takes place within the nanoscale free-electron model, a structural and dynamical model of nanowires that treats the electron-confinement effects exactly while replacing the atomic structure by a continuum.
Electron-shell effects, previously shown to be responsible for the stability of wires with magic radii, favor the formation of kinks connecting magic cylinders.
A rich kink dynamics including interkink interactions ensues and is similar to that observed experimentally.
\end{abstract}


\maketitle \vskip2pc

\label{sec:intro}

The formation and thinning dynamics of metallic nanowires suspended between macroscopic contacts has recently been imaged, in real time and with atomic resolution, using transmission electron microscopy (TEM).\cite{kondo97,kondo00,rodrigues00,rodrigues02,oshima03}
Almost perfect Au (Refs.\ \onlinecite{kondo97}, \onlinecite{kondo00}, and \onlinecite{oshima03})  and Ag (Refs.\ \onlinecite{rodrigues00}, and \onlinecite{rodrigues02}) cylindrical wires, with diameters ranging\cite{kondo00} from 5 to $15$ \AA, are found to form and be stable for seconds at room temperature.
These cylinders usually do not break up at once but are instead seen to thin down one step at a time by an amount close to one atomic layer.\cite{oshima03}
Real-time movies\cite{TakayanagiMovies} of the thinning dynamics reveal that wires can remain in a metastable state for a time of order of a second, until a kinklike structure nucleates at one end of the wire and subsequently moves along the cylinder.
The kink formation and initial displacement over a significant portion of the wire are extremely rapid, of the order of the time resolution of the experiment, while subsequent motion is slower and takes a significant fraction of a second\cite{oshima03} for a displacement of order of $1$ nm.
In some cases, a kink appears to stop along the wire until it is joined by another one, at which point they move along together until they are absorbed by the contact at the other end of the wire.

Due to the large fraction of surface atoms---with low coordination numbers---in such nanowires, surface effects are particularly important:
Barring an additional stabilizing mechanism, they are expected to trigger a Rayleigh instability\cite{chandrasekhar81,kassubek01} that would break the wire up.
Surface tension indeed seems a likely candidate for the force driving nanowire thinning but cannot account for the step by step nature of the process.
A layer by layer thinning of a crystalline structure could, in principle, account for it,\cite{jagla01} but comparison of TEM images with simulations\cite{kondo00,oshima03a} leads to the conclusion that such thin nanowires do not have a crystalline structure.
They adopt instead a multishell structure with cylindrical symmetry, similar to carbon nanotubes, which would not be classically stable.
This suggests that the atoms adjust themselves as closely as possible to the electronic structure, which favors cylindrical symmetry,\cite{urban06} which cannot be done while keeping a crystalline structure.

Electron-shell effects---similar to those well-known in cluster physics \cite{brack93}---have been shown to be essential to the wire stability,\cite{kassubek01,zhang03,burki05} providing an energy barrier sufficient to make nanowires metastable, with lifetimes\cite{burki05a} of order of a second.
This energy barrier arises from the competition between surface tension and electron shell effects, the latter
resulting from the quantum confinement of conduction electrons within the cross section of the wire.
The ``jerkiness'' of the thinning suggests a thermally activated process, where random fluctuations are occasionally large enough to overcome the barrier, thus triggering thinning to the next metastable wire.\cite{burki05a}
This picture is reinforced by the fact that thinning forced by electron irradiation of the wire follows the same mechanism, but is much more regular.\cite{zandbergen05}

Any meaningful theoretical description of this problem requires a model that captures both surface and electron-shell effects.
In addition, it needs to be simple enough to allow simulations of large systems over long time scales.
Both of these conditions are fulfilled by the nanoscale free-electron model (NFEM),\cite{stafford97,burki03,burki05} a continuum model where the atomic structure is replaced by a uniform, positively charged background, and the emphasis is put on the electronic structure.
{\sl Ab initio} methods, which include both electronic and ionic structures, satisfy the first of the above conditions, but not the second.
They have been used to confirm the atomistic shell structure of gold nanowires through energy minimization,\cite{tosatti01} but they are computationally quite intensive and limited to relatively thin wires even for structure calculations.
As the observed thinning dynamics involves hundreds, if not thousands, of atoms and takes place over a sizable fraction of a second, it is far beyond the reach of {\sl ab initio} simulations,\cite{tosatti01,dasilva04} which are typically limited to a few dozen atoms and cover a nanosecond.
Even classical molecular dynamics,\cite{jagla01,koh06} which do not include electron-shell effects, can only simulate a few nanoseconds at most.

The NFEM has successfully explained the linear stability of cylindrical nanowires with ``magic'' radii as resulting from a competition between surface and electron-shell effects.\cite{kassubek01,zhang03}
As a continuum model, the NFEM can obviously not address issues such as the influence of the crystalline structure of the macroscopic contacts on the wire stability.\cite{jagla01}
The model has proved useful in understanding ionic dynamics through surface self-diffusion over long time scales.\cite{burki03}
In particular, a random wire has been shown to naturally evolve into a universal equilibrium shape consisting of a perfect cylinder of a magic radius, connected to thicker leads through an abrupt junction.\cite{burki03,burki04}
A stochastic model of the dynamics of a nanocylinder under thermal fluctuations\cite{burki05a} has predicted long lifetimes compatible with experimental observations.
The escape mechanism for long wires has been shown to consist of the nucleation of a kink at one end of the wire.
The kink subsequently propagates along the wire, leaving behind a thinner or thicker cylinder.
This suggests that a dynamics similar to experiments may take place within the NFEM.

Motivated by these results, I argue that, despite being a fully continuum model, the NFEM contains an intrinsic length scale---the Fermi wavelength $\lambda_F$---which, through electron-confinement effects, provides some degree of ``discreteness'' for the ionic dynamics.
The competition between surface and electron-shell effects favors the formation of kinks, \cite{burki05a} or solitons, connecting cylinders of magic radii.
The dynamics of these solitons is shown to be in semiquantitative agreement with experiments in many details.
In particular, the model predicts an extremely fast nucleation and initial propagation of kinks at the wire end, followed by a regular, slower motion.
Interactions between kinks, which can be either attractive or repulsive, profoundly affect their dynamics.
For some wires, a soliton is seen to be essentially stopped by its interaction with another one, until they combine and move as a single, larger kink, much like what is observed experimentally.\cite{oshima03,TakayanagiMovies}

The paper is organized as follows. In Sec.\ \ref{sec:model}, the NFEM is introduced and some of its main results directly relevant to the present article are summarized. 
The formation and propagation of a soliton are discussed in Sec.\ \ref{sec:singlekink}, while interactions between multiple kinks are analyzed in Sec.\ \ref{sec:multkinks}.
A discussion of the results and their relation to existing experiments is provided in the last section.

\section{The nanoscale free-electron model}\label{sec:model}

The NFEM is a continuum model of open metallic nanosystems with emphasis on the electronic structure, which is treated exactly.\cite{stafford97}
It is thus particularly suitable as a model of metal nanowires, where electron-confinement effects have been shown to influence both transport and cohesive properties\cite{rubio96,brom99} (see Ref.\ \onlinecite{agrait02} for a recent review.)
The discrete atomic structure is replaced by a uniform, positively charged background (jellium), which provides a confining potential for electrons.
Electronic degrees of freedom are described using a free-electron model, thus neglecting interactions, except inasmuch as they rescale\cite{kassubek99,zhang03,burki05} macroscopic quantities such as the bulk energy density $\omega_B$, and the surface tension $\sigma_s$.

This model is particularly suitable for simple metals, with good screening and a close-to-spherical Fermi surface, such as those with a single $s$-electron conduction band at the Fermi surface.
Such conditions are fulfilled for alkali metals such as sodium and for noble metals such as gold and silver, although $d$ electrons may play a role for noble metals.
The NFEM has provided an understanding in simple physical terms of many transport\cite{stafford97,kassubek99,burki99,burki99a}, stability,\cite{kassubek01,zhang03,urban04,urban06} and dynamical\cite{burki03,burki04,burki05a,burki05} properties of alkali and noble metal nanowires, in quantitative agreement with experiments..

While the NFEM has no such restriction, only axisymmetric wires are considered in this paper.
This choice is justified by the facts that 
(i) the most stable wires are axisymmetric,\cite{urban04,urban06} with only a small fraction of low conductance wires having a broken axial symmetry, and
(ii) the dynamics tends to decrease surface area and thus further favors axisymmetric wires.
In addition, axial symmetry greatly simplifies the numerical treatment of the dynamics, as the wire shape can be described by a single radius function $R(z,t)$.

A nanowire connected to macroscopic contacts being an open system, the electronic energy is given by the grand canonical potential $\Omega_e$.
Like any extensive thermodynamic quantity, $\Omega_e$ can be written as a Weyl expansion\cite{brack97}---a series in geometrical quantities such as system volume ${\cal V}$ and surface area ${\cal S}$---complemented by a mesoscopic, fluctuating contribution $\delta\Omega$:
\begin{equation}\label{eq:Omega_e}
  \Omega_e[R(z)] = \omega_B{\cal V} + \sigma_s{\cal S} +\delta\Omega,
\end{equation}
where the values of $\omega_B$ and $\sigma_s$ may be chosen to match the bulk properties of the metal to be described.
As results are independent of $\omega_B$, its free-electron value $\omega_B=-2\Ef k_F^3/15\pi^2$ is used, while the surface tension is set to $\sigma_s=1.256\,$N/m, a value appropriate for the description of gold.\cite{tyson77}

\begin{figure}[t]
  \includegraphics[width=0.99\columnwidth]{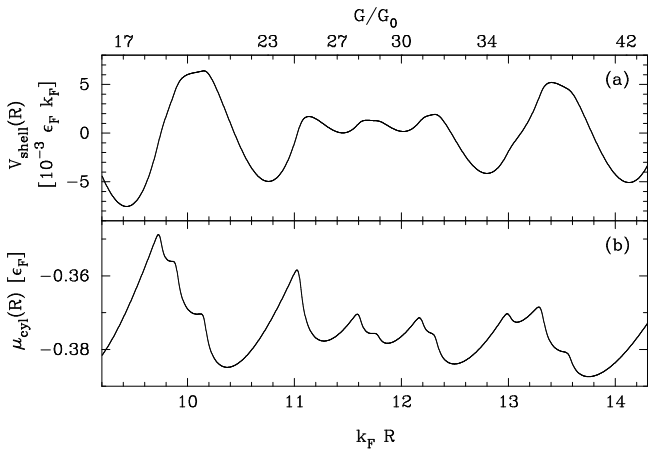}
  \caption{(a) Electron-shell potential $V_{shell}(R)$, and 
        (b) atomic chemical potential $\mu_{cyl}(R)$ for a cylindrical wire,
        as a function of its dimensionless radius $k_FR$, where $k_F$ is the Fermi wavevector.
        (The radius range is limited to that corresponding to the simulations presented here. 
        For a graph over a more extended range, see Ref.\ \onlinecite{burki05}.)
  	The top axis shows the conductance values of linearly stable cylinders in units of the 
  	conductance quantum, $G_0=2e^2/h$.
  }\label{fig:Vshell}
\end{figure}

Assuming the wire cross section varies slowly along the wire (adiabatic approximation), the mesoscopic contribution, which is of purely quantum-mechanical origin, may be written as 
\begin{equation}\label{eq:dOmega}
 \delta\Omega[R(z)] = \int_0^L\!dz\,V_{shell}[R(z)],
\end{equation}
where the electron-shell potential $V_{shell}(R)$ can be computed using a semi-classical approximation.\cite{burki05}
The name ``electron-shell potential'' refers to the analogous shell closing observed in metal clusters.\cite{brack93}
Although there is strictly speaking no shell closing in metal nanowires,\cite{tosatti01} since they are open systems, a large gap in the transverse eigenenergies at the Fermi energy increases the stability\cite{urban04} of the corresponding nanowire.
This results in a deep minimum of $V_{shell}(R)$, depicted in Fig.\ \ref{fig:Vshell}(a) for a cylindrical wire as a function of its radius $R$.
A linear stability analysis\cite{kassubek01,zhang03,urban03} shows that wires within a finite interval around the magic radii are stable toward all small perturbations.

The ionic dynamics is taken to be classical and can be assumed to occur mainly through surface self-diffusion, as most atoms in thin metal wires are surface atoms.\cite{SurfAtoms,burki03}
The evolution equation for the radius function $R(z,t)$ derives from ionic mass conservation
\begin{equation}\label{eq:diffusion}  
  \frac{\pi}{{\cal V}_a}\frac{\partial R^2(z,t)}{\partial t}
        + \frac{\partial}{\partial z}J_z(z,t)
	= 0,
\end{equation}  
where ${\cal V}_a=3\pi^2/k_F^3$ is the volume of an atom in a monovalent metal, and the $z$ component $J_z$ of the total surface current is given by Fick's law:
\begin{equation}\label{eq:current} 
  J_z = -\frac{\rho_SD_S}{k_B T}\frac{2\pi R(z,t)}{\sqrt{1+(\partial_zR)^2}}
        \frac{\partial\mu}{\partial z}.
\end{equation}  
Here, $\rho_S$ and $D_S$ are, respectively, the surface density of ions and
the surface self-diffusion coefficient, and $\partial_zR=\partial R/\partial z$.

The ionic chemical potential $\mu[R(z)]$ can be computed from the energy change due to the local addition of the volume ${\cal V}_a$ of an atom to the system.
Within the Born-Oppenheimer approximation, and assuming that the electrons act as an incompressible fluid,\cite{zhang03,burki05,incompFluid} the chemical potential is given\cite{burki03} by the functional derivative $\mu[R(z)]=[{\cal V}_a/(2\pi R)]\times[\delta\Omega_e/\delta R(z)]$. 
Starting from Eqs. (\ref{eq:Omega_e}) and (\ref{eq:dOmega}), one obtains
\begin{equation}\label{eq:mu}  
  \mu[R(z)] = \mu_0
    + \frac{{\cal V}_a}{2\pi R}\left(
      \frac{2\sigma_s\partial{\cal C}[R(z)]}{\sqrt{1+(\partial_z R)^2}}
      + \frac{\partial V_{shell}}{\partial R}\right),
\end{equation}  
where $\mu_0=\omega_B {\cal V}_a$ is the bulk chemical potential.
Here, $\partial{\cal C}[R(z)] = \pi\left(1 - \frac{R \, \partial^2_{z}R}{1+(\partial_zR)^2}\right)$ is the local mean curvature of the wire and results from the functional derivative of the surface term.
The chemical potential of a cylinder $\mu_{cyl}(R)\equiv\mu[R(z)=R]$ is plotted as a function of the radius $R$ in Fig.\ \ref{fig:Vshell}(b).

The precise value of $D_S$ in Eq.\ (\ref{eq:current}) is not known for most metals, but it can be removed from the evolution equation by rescaling time to the dimensionless variable $\tau=\omega_0t$, with the characteristic temperature-dependent frequency $\omega_0=\rho_SD_ST_F/T$.
For comparison with experimental time scales, one can estimate that for quasi-one-dimensional diffusion, $D_s \approx \nu_D a^2 \exp(-E_s/k_B T)$,
where $\nu_D$ is the Debye frequency, $a$ is the lattice spacing, and $E_s$ is an
activation energy comparable to the energy of a single bond in the solid.

As mentioned above, a linear stability analysis\cite{kassubek01} finds intervals of stable radii.
The criterion for linear stability of a cylindrical wire has been shown\cite{burki05} to be
\begin{equation}\label{eq:stab_mu}
  \frac{d\mu_{cyl}(R)}{dR} > 0.
\end{equation}
A linearized dynamical theory\cite{zhang03} suggests that unstable wires develop an exponentially growing instability with a well-defined wavelength corresponding to the maximally unstable mode.
This instability was argued to saturate and eventually lead to a phase separation of the wire into thick and thin segments of stable radii. 
Simulations using the full dynamics, defined by Eqs.\ (\ref{eq:diffusion})--(\ref{eq:mu}), have confirmed this and have shown that the phase separation occurs via a complex dynamics involving kink interactions and annihilation.\cite{burki03}

Note that, despite the apparent simplicity of the evolution equations [Eqs.\ (\ref{eq:diffusion})--(\ref{eq:mu})] as written above, the resulting partial differential equation for $R(z,t)$ is fourth order in $z$ derivatives and highly nonlinear.
Its classical counterpart, with $V_{shell}\equiv 0$, has been extensively studied.\cite{bernoff98}
It was shown to have stationary states corresponding to shapes of constant mean curvature, the sphere, the cylinder, and the unduloid of revolution, the latter being always unstable.
The addition of the electron-shell potential, of quantum-mechanical origin, stabilizes magic cylinders connected to unduloid-shaped leads as the universal equilibrium shape\cite{burki03,burki04} and leads to the rich kink dynamics, as discussed in this paper.

Given the universality of the equilibrium shape and the ``sharpness'' of the connection between the unduloid-shaped leads and the cylindrical wire,\cite{burki03,burki04} one can focus on the dynamics of the cylindrical part of the wire, assuming that the lead is large enough to serve as a source or sink of atoms for the wire without undergoing significant changes.
In this case, Neumann boundary conditions, $\partial_z R = 0$ at both ends of the wire, $z=0$ and $z=L$, are found to best describe the connection to the leads. 
The diffusive dynamics being volume conserving, a sink or source of atoms at one or both wire ends must be explicitly added to mimic a nonequilibrium situation, where an ionic current flows between the cylinder and the leads, thus allowing for thinning or growth of the wire. 
This is done by adding a boundary current term to the right-hand side of Eq.\ (\ref{eq:diffusion}):
\begin{equation}\label{eq:sink}
 J_{bnd} =  J_{l}\delta(z) - J_{r}\delta(z-L),
\end{equation}
where a positive $J_{l(r)}$ is a source of atoms at the left (right) wire end, while negative values correspond to sinks of atoms.

\section{Single kink dynamics}\label{sec:singlekink}

\begin{figure}[t]
 \centering
 \includegraphics[width=0.99\columnwidth]{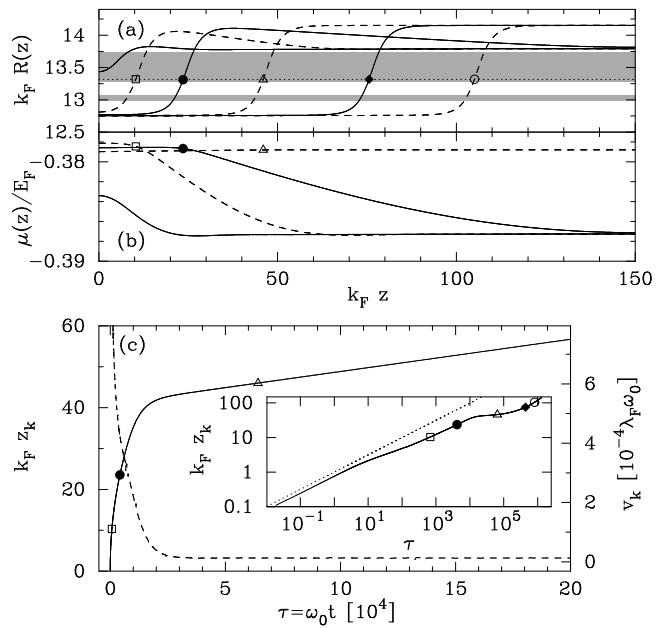}
 \caption{(a) Radius function $R(z)$ at various evolution times during the transition 
          $G/G_0=42\rightarrow34$ for a wire subject to a sink of atoms at $z=0$.
          Solid and dashed lines are used on alternate curves for clarity, 
          while the gray areas correspond to unstable radii, and the dotted line shows 
          the radius used to define the kink position $z_k$.
          (b) Corresponding chemical potential profiles $\mu(z)$ for the first four curves in (a).
          The last one looks constant on this scale, but has a slight positive slope left of the kink.
          (c) Kink position $z_k$ (solid line) as a function of time on linear and log-log (inset) scales.
          The kink position for the shapes plotted in the top panel are marked with corresponding 
          symbols on each graph.
          The dashed line gives the kink velocity $v_k$, with units given on the right axis.
 }\label{fig:kinkpos}
\end{figure}

Motivated by the experimental observation of wire thinning,\cite{oshima03} I will consider mainly the case of a single sink of atoms at the left end of the wire, $J_l < 0$ and $J_r = 0$.
A number of simulations for various wire lengths and boundary currents have been performed.
Figures\ \ref{fig:kinkpos}, \ref{fig:kinkposition}, and \ref{fig:kinkinteract} show results for $k_FL=150$, or $L\simeq12$ nm for Au, with a boundary current $J_{l}=-10^{-4}\pi\omega_0$.
The parameter dependence of the results, as well as the influence of a second sink of atoms, or fluctuations of the boundary current, are discussed in Sec.\ \ref{sec:concl}.

A boundary current $J_l<0$ creates, following Eq.\ (\ref{eq:current}), a positive gradient of chemical potential $\mu(z)$ that progressively expands along the wire.
From Eq.\ (\ref{eq:stab_mu}), a stable wire corresponds to a positive slope of $\mu_{cyl}(R)$.
Thus, a positive gradient of $R(z)$ corresponds to the gradient of $\mu(z)$.
Once the radius change at the boundary is large enough to cross a stability threshold, corresponding to a minimum of $\mu_{cyl}(R)$, an instability sets in\cite{zhang03,burki03} and leads to the formation of a soliton that will propagate along the wire.
In most cases, like the transition from $G/G_0=42\rightarrow34$ (Fig.\ \ref{fig:kinkpos}), or $G/G_0=23\rightarrow17$ ($G_0=2e^2/h$ is the quantum of conductance), a single kink nucleates and propagates along the whole wire.
The present section discusses this case in detail.
Sometimes, however, a second kink nucleates before the first one has reached the end of the wire, as is the case in the transition $G/G_0=34\rightarrow27\rightarrow23$ (Fig.\ \ref{fig:kinkposition}.) 
If this happens, the two solitons interact attractively and eventually combine and propagate as a single, larger kink (Figs.\ \ref{fig:kinkposition} and \ref{fig:kinkinteract}.)
This is the simplest case of kink interactions, which are discussed in Sec.\ \ref{sec:multkinks}.

Figure\ \ref{fig:kinkpos}(a) displays the radius function $R(z,\tau)$ at different evolution times (with alternating solid and dashed lines for clarity), with a dotted horizontal line showing the channel opening threshold,\cite{stafford97} used to track the kink position $z_k$.
The radii corresponding to unstable cylinders are emphasized with a gray background.
The corresponding chemical potential profiles $\mu(z,\tau)$ for the first four configurations are shown in Fig.\ \ref{fig:kinkpos}(b).
(The last two configurations are omitted as they are indistinguishable from the last displayed one, which looks constant on the scale used.)
The lower panel (c) shows $z_k$ as a function of time (solid line), both on linear (main plot) and log-log (inset) scales, as well as the kink velocity $v_k$ (dashed line).
One can clearly identify two phases with different time evolutions of the kink position: the kink formation, during which $z_k$ increases quickly but the kink velocity $v_k$ decreases abruptly, and the kink propagation, when the kink moves without deformation at a constant velocity $v_0$. The kink formation actually takes place in two stages, as will be discussed shortly.

As soon as the wire radius at the sink of atoms reaches a critical radius $R_c$, corresponding to a marginally stable wire [upper limit of gray area in Fig.\ \ref{fig:kinkpos}(a)], for which $d\mu_{cyl}/dR(R_c)=0$ (first curve from the left), an instability sets in\cite{zhang03} and grows exponentially, leading to the quick formation of a kink (second curve from the left).
During this first stage of the kink formation, the radius ``overshoots'' ahead of the kink and creates a ``bump'' visible in the first dashed curve.
In a second stage, this overshoot progressively spreads over the whole wire (second to fourth curves from the left), dragging the kink behind itself at a quickly reducing speed.
The kink subsequently propagates without deformation (remaining curves) with a constant velocity, proportional to the boundary current $|J_{l}|$, provided the wire is stable enough, not too long, and the current small enough (see Sec.\ \ref{sec:multkinks}).

\subsection{Kink formation}\label{par:KinkFormation}
When the wire radius at the boundary reaches an instability threshold $R_c$, an instability with a single wavelength $\lambda=2\pi/q$ starts growing exponentially at a rate $\omega$,\cite{zhang03} as discussed in Sec.\ \ref{sec:model}.
As the instability is driven by the boundary current, it is initially localized at and propagates from the wire end. This is accounted for by a propagating envelope function chosen as an exponential with decay length $\xi$ and propagation speed $v$, which multiplies the perturbation function $\delta R(z)$, which can thus be written as
\begin{equation}
 \delta R\sim \text{Re}\left\{\exp[i(\omega t-q z)]\exp[-(z-v t)/\xi]\right\}.
\end{equation}
Although there is no well-defined kink at this point, its position can be defined as the point where $\delta R$ reaches a fixed value $\delta$. 
Results depend weakly on that value,\cite{vansaarloos03} which is chosen to be the threshold for conduction-band closing. 

As happens for the instability of a cylinder,\cite{zhang03,burki03} the growth saturates, leaving a well-formed kink, once it reaches a zone of stable radii, i.e., grows out of the gray areas in Fig.\ \ref{fig:kinkpos}(a). 
In this particular case, the instability grows through two unstable zones, shown by the two gray areas, as the intermediate radii are not stable enough to stop its exponential growth.

The product of the decay length $\xi$ of the envelope and wave vector $q$ of the perturbation can be shown,\cite{burki07b} using the linearized version of evolution equations [Eqs.\ (\ref{eq:diffusion})--(\ref{eq:mu})], to be a constant $q\xi\simeq 3.2$, so that the instability is only visible over essentially one wavelength $\lambda$.
Its first half wavelength becomes the kink, while the second half wavelength, much reduced in amplitude, is responsible for the observed overshoot.

Solving $\delta R(z_k, t)\equiv\delta$ for small $q z_k(t)$, neglecting the envelope at first, the kink position during its formation is found to be
\begin{equation}\label{eq:zk_formation}
  k_F z_k(\tau) \propto \sqrt{\omega_0(t-t_0)},
\end{equation}
where $t_0$ is the kink formation time.
This result is independent of the boundary current, as long as the latter is sufficiently small compared to the growth rate of the instability. 
Corrections to this behavior due to the envelope are small in its range of validity, as its propagation speed is much smaller than its growth rate.
This $t^{1/2}$ behavior of the kink position is observed only in the early stages of the kink formation, $\Delta\tau=\omega_0(t-t_0)\lesssim2\times10^4$, as can be seen in the inset of Fig.\ \ref{fig:kinkpos}, where $\sqrt{\tau}$ has been plotted as a dotted line for comparison.

\subsection{Overshoot spreading}\label{par:Overshoot}

The progressive spreading of the overshoot over the whole wire, driven by its far-from-equilibrium chemical potential [rightmost solid line in Fig.\ \ref{fig:kinkpos}(b)], speeds up the kink propagation, as is discussed in this section.

Prior to the kink formation, the wire radius is driven by the boundary current out of its most stable value, which corresponds to a minimum of $V_{shell}(R)$.
It is instead close to its marginally stable value which, through Eq.\ (\ref{eq:stab_mu}), corresponds to a minimum of $\mu_{cyl}(R)$.
The overshoot brings a portion of the wire back to its most stable radius $R_r$, creating a negative gradient of chemical potential and, according to Eq.\ (\ref{eq:current}), a corresponding ionic current in the forward direction.
This current favors the spreading of the overshoot and drags the kink along with it.

\begin{figure}[t]
  \includegraphics[width=0.99\columnwidth]{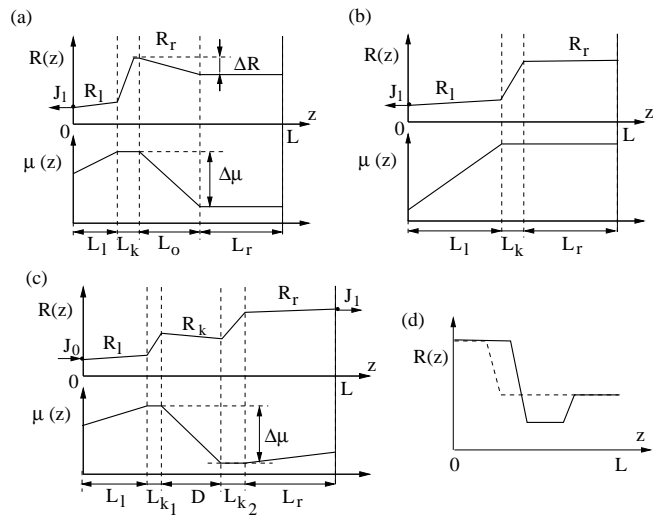}
  \caption{Schematic diagram of the radius function $R(z)$, and corresponding chemical potential $\mu(z)$, 
           during (a) the overshoot propagation ahead of a kink, (b) a soliton propagation, 
           (c) interaction of two kinks, and (d) interaction of a kink and antikink.
           The diagrams define the radii and lengths used in the text, as well as the chemical 
           potential drop $\Delta\mu$. 
           The arrows labeled $J_{\kappa}$ mark the positions of sinks or sources of atoms.
           In panel (d), only the radius function is shown, and the dashed line represents $R(z)$ before 
           nucleation of the upward kink.
  }\label{fig:kinkSchema}
\end{figure}

The chemical potential along the wire during overshoot propagation is depicted schematically in Fig.\ \ref{fig:kinkSchema}(a).
It has a constant slope, corresponding to a slight gradient of $R(z)$ (see beginning of the present section), proportional to the boundary current on the left of the kink and is constant in the cylindrical part of the wire on the right of the overshoot.
It is essentially constant over the kink and drops by an amount $\Delta\mu<0$ over the extent of the overshoot. 
As discussed above, $\Delta\mu$ is essentially fixed by the potential difference between the most stable and the corresponding marginally stable wires in the stability interval considered, and is thus constant during evolution.
Consequently, the gradient of potential $\Delta\mu/L_o$ over the overshoot, of length $L_o$,  decreases (in absolute value) with time. 

With these elements in mind, the overshoot spreading, and the corresponding kink propagation, can be fully understood from the fact that the evolution [Eq.\ (\ref{eq:diffusion})] is volume conserving.
Let us consider a portion of the wire of length $L_0>L_o$, starting at the overshoot maximum.
Approximating the overshoot as a cone of average radius $R_r-\Delta R/2$ [see Fig.\ \ref{fig:kinkSchema}(a)], the volume $V_0=\pi[(R_r-\Delta R/2)^2L_o+(R_r-\Delta R)^2(L_0-L_o)]$ of this section of the wire increases over time due to an incoming current $J=-2\pi\omega_0R_r\Delta\mu/L_o$ on its left-hand side, while there is no current on its right-hand side, which forces the overshoot expansion.
Integrating Eq.\ (\ref{eq:diffusion}) along this portion of the wire, one obtains the following mass conservation equation:
\begin{equation}\label{eq:MassConservation}
  \frac1{{\cal V}_a}\frac{dV_0}{dt}=J_{left}-J_{right}=J,
\end{equation}
where $J_{left(right)}$ are the ionic currents on the left-hand(right-hand) side of the section of the wire considered; in this case, their values are $J$ and $0$, respectively.
This equation yields the time dependence of the overshoot length, $L_o(\Delta\tau)=l_o\sqrt{\Delta\tau}$, where $\Delta\tau$ measures the time from its formation, and
\begin{equation}\label{eq:Lbump}
  l_o^2=-\frac{4{\cal V}_aR_r\Delta\mu}{\Ef\Delta R(R_r-3\Delta R/4)}.
\end{equation}
The parameters in Eq.\ (\ref{eq:Lbump}), $R_r$, $\Delta R$, and $\Delta\mu=\mu_{cyl}(R_l)-\mu_{cyl}(R_r-\Delta R)$, can all be obtained from the shape $R(z,\tau)$ at any given time during overshoot propagation and determine its expansion.

The time evolution of the kink displacement $\Delta z_k(\Delta\tau=\tau-\tau_1)\equiv z_k(\tau)-z_k(\tau_1)$, $z_k$ being the kink position, during the overshoot propagation, derives from mass conservation [Eq.\ (\ref{eq:MassConservation})] for the whole wire and is
\begin{equation}\label{eq:zk_bump}
  \Delta z_k(\Delta\tau) = 
    \frac{\Delta R(R_r-\frac{3\Delta R}4)l_o\Delta\tau^{1/2} - \frac{{\cal V}_aJ_l}{\pi\omega_0}\Delta\tau}
          {(R_r-\Delta R)^2-R_l^2}.
\end{equation}

In the present simulation, $J_{l}/\omega_0\ll1$, so that the corresponding term in Eq.\ (\ref{eq:zk_bump}) may be neglected and only the $\sqrt{\Delta\tau}$ behavior is observable, as can be seen in the inset of Fig.\ \ref{fig:kinkpos}(c) for $10^2\lesssim\tau\lesssim10^4$.

How far the kink moves during the overshoot expansion, as well as how long it takes for the overshoot to reach the wire end, can be estimated from $\Delta z_k(\Delta\tau_{o})+L_o(\Delta\tau_{o})\sim L$. 
It is found that $\Delta z_k(\Delta\tau_{o})\simeq \eta L$, with a proportionality constant $\eta$ depending on the coefficients of $\Delta\tau$ and $\sqrt{\Delta\tau}$ in Eq.\ (\ref{eq:zk_bump}). 
For most kinks, $\eta\sim0.1-0.3$. 
The propagation time $\Delta\tau_{o}$ is also proportional to $L$ and is found to be of order $10^3-10^4$.
For the kink of Fig.\ \ref{fig:kinkpos}, the constant is
$\eta\simeq0.14$, so that the kink jumps ahead by about $k_F\Delta L\simeq20$ during this phase that lasts a time $\Delta\tau_{o}\simeq4\cdot10^3$, in agreement with the simulation.

\subsection{Kink propagation}\label{par:KinkProp}

Once the overshoot has reached the end of the wire, the chemical potential profile becomes simpler and is then constant for $z>z_k(\tau)$, as depicted schematically in Fig.\ \ref{fig:kinkSchema}(b). 
The same volume conservation argument, applied to the whole wire, shows that the kink moves with a constant speed determined by the radii on both ends of the wire and proportional to the boundary current:
\begin{equation}\label{eq:zk_prop}
  z_k(\tau) = z_k(\tau_1) - \frac{{\cal V}_a}{\pi(R_r^2-R_l^2)}\frac{J_l}{\omega_0}(\tau-\tau_1),
\end{equation}
where $\tau_1$ is the time at which the overshoot has been fully absorbed at the wire end.
The kink velocity obtained from Eq.\ (\ref{eq:zk_prop}) is in good agreement with that of Fig.\ \ref{fig:kinkpos}(c).

None of this discussion depends critically on the sign of $J_l$, and the same kind of soliton formation and propagation is observed for a source of atoms.

\section{Kink interactions}\label{sec:multkinks}
%

The likelihood of a second kink being nucleated before the first reaches the wire end increases with the boundary current $|J_{l}|$ and wire length $L$ but decreases with the wire stability.
As mentioned previously, the chemical potential profile on the left of the kink has a constant gradient, proportional to $J_l$, corresponding to a small gradient of $R(z)$. 
Inevitably, if the kink were propagating into an infinite wire, there would be a time $\tau_n$ when the kink position $z_k(\tau_n)$ is such that the potential difference due to this gradient is large enough for a new instability to set in, and thus for a new kink to nucleate. 
If the wire length $L$ is larger than $L_c\equiv z_k(\tau_n)$, this will happen even for a finite wire.
The critical length $L_c$ increases with the wire stability: A given drop $\Delta\mu$ of the chemical potential may be enough to trigger nucleation of a new kink for a given wire, say, one with $G=27\,G_0$, but not for a more stable wire, such as one with $G=34\,G_0$ (see Fig.\ \ref{fig:Vshell}). As the gradient of chemical potential is proportional to the boundary current, according to Eq.\ (\ref{eq:current}), one clearly has $L_c\propto |J_l|$.

\begin{figure}[b]
 \centering
 \includegraphics[width=0.99\columnwidth]{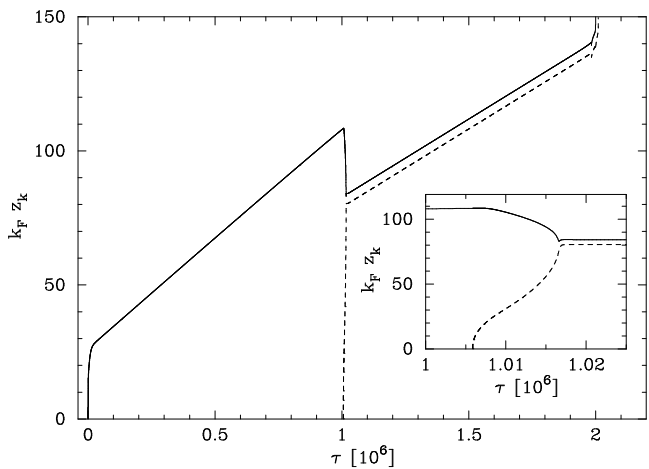}
 \caption{Evolution of the kink positions for two interacting kinks in a wire subjected 
          to a sink of atoms at its left boundary.
          The wire radius before nucleation of the first kink ($\tau < 0$) is $k_FR=12.75$, 
          corresponding to a conductance $G=34G_0$, and the wire thins down to $k_FR=10.7$, 
          and $G=23G_0$, by the time the `combined' kink has reached the wire's end.
          The inset zooms in on the time region where the two kinks interact ($\tau\simeq10^6$).
 }\label{fig:kinkposition}
\end{figure}
\begin{figure}[t]
 \centering
   \includegraphics[width=0.99\columnwidth]{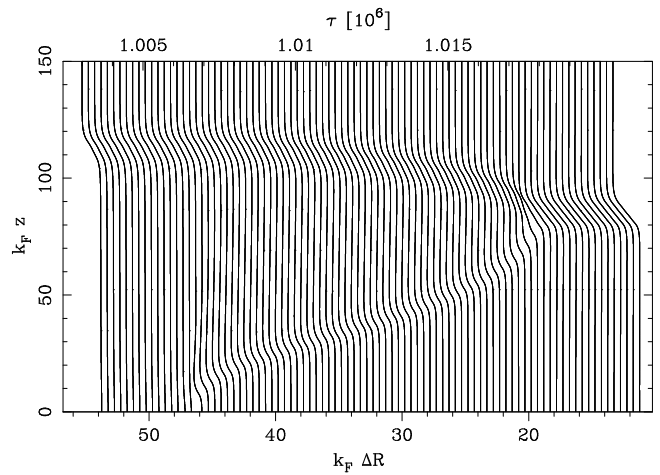}
 \caption{Radius function $R(z)$ at equidistant times $\tau$, showing the nucleation of a second kink 
          (dashed line in Fig.\ \ref{fig:kinkposition}) that interacts and fuses with the preexisting 
          kink around $\tau\simeq1.016\cdot10^6$. 
          $R(z)$ is plotted vertically, with a rightward shift proportional to the evolution time, 
          for better comparison with the inset of Fig.\ \ref{fig:kinkposition}.
 }\label{fig:kinkinteract}
\end{figure}

During the transition from $G/G_0=34$ to $23$, via $27$, $L_c$ happens to be shorter than the wire length in the simulation discussed above, so that a second soliton is nucleated.
Figure \ref{fig:kinkposition} traces the positions of the two interacting kinks. 
Under the influence of the newly nucleated kink, appearing around $\tau=10^6$ (dashed line), the existing kink slows down and even reverses its motion (solid line). 
The approach speed of the two kinks is observed to increase as they get closer (see inset).
After they interact, both solitons appear to move in parallel at a constant speed. 
In fact, following the radius function $R(z,\tau)$ during evolution, plotted vertically with a rightward shift proportional to $\tau$ in Fig.\ \ref{fig:kinkinteract}, reveals that the two kinks combine to form a larger one.

The interaction between any two kinks can be understood in simple terms when both are well formed and separated enough, i.e., each of them propagates without deformation and all overshoot expansion is over. 
In that case, the kinks connect approximately cylindrical segments of the wire (each with a slight slope corresponding to a small ionic current) with respective radii of $R_l$, $R_k$, and $R_r$, as depicted schematically in Fig.\ \ref{fig:kinkSchema}(c).
According to Eq.\ (\ref{eq:current}), the chemical potential profile on either side of the pair of solitons has slopes proportional to the currents $J_0$ and $J_1$ on the left- and right-hand sides, respectively, which are assumed to be known.
There is, in general, a linear change of the potential $\Delta\mu=\mu_{cyl}(R_l)-\mu_{cyl}(R_r)$ over the segment of the wire of length $D$ joining the two kinks.
Considering mass conservation [Eq.\ (\ref{eq:MassConservation})] separately for two portions of wire containing, respectively, only the first and second solitons, and combining the two equations thus obtained, the following differential equation for the distance $D$ between the kinks is obtained:
\begin{multline}\label{eq:kinkdist}
  \frac{\pi}{{\cal V}_a}\frac{dD(t)}{dt}=\frac{J_0}{R_k^2-R_l^2} + \frac{J_1}{R_r^2-R_k^2} \\
    + \frac{2\pi\omega_0}{\Ef}
      \frac{R_k(R_r^2-R_l^2)}{(R_r^2-R_k^2)(R_k^2-R_l^2)}\frac{\Delta\mu}{D(t)}.
\end{multline}

As long as the outside currents $J_0$ and $J_1$ are small enough, the last term in Eq.\ (\ref{eq:kinkdist}) dominates, so that $D^2(t)\propto(t-t_0)$.
The proportionality constant can have either sign depending on the balance of chemical potentials of the outside cylinders, as well as the type of both kinks (upward or downward solitons).
The interaction force between kinks can thus be attractive or repulsive.
It can be calculated as $F=-\left.\frac{\partial\Omega_e}{\partial D}\right|_{\cal V}$, where $\Omega_e$, given by Eq.\ (\ref{eq:Omega_e}), is the energy of a portion of the wire containing both kinks.
The result is
\begin{equation}\label{eq:Fkink}
  F = -\varepsilon(R_k) 
      + \frac{R_r^2-R_k^2}{R_r^2-R_l^2}\varepsilon(R_l)
      + \frac{R_k^2-R_l^2}{R_r^2-R_l^2}\varepsilon(R_r),
\end{equation}
where $\varepsilon(R)=2\pi\sigma_sR+V_{shell}(R)$ is the linear energy density for a cylinder of radius $R$.
The interkink force is thus {\sl constant}, independent of their separation.

Coming back to the solitons of Fig.\ \ref{fig:kinkposition}, their interaction is found to be attractive, as expected from their observed behavior.
Even for the largest kink separation, of order $k_FD\sim100$, the last term of Eq.\ (\ref{eq:kinkdist}) is still one order of magnitude larger than the other terms, which may thus be neglected. 
After solving for $D(t)$, the time evolution for the kink positions $z_{k_1}(\tau)$ and $z_{k_2}(\tau)$ can be extracted by considering volume conservation for the whole wire.
Results obtained through this simplified evolution are found to be in agreement with the full simulation.

A simple case of repulsive interaction between two kinks is obtained by replacing the sink of atoms by a source, $J_l>0$.
While the slightly stable wire at $G=30\,G_0$ was skipped during wire thinning, or more exactly a new kink was nucleated so quickly that that radius was only observable for a very short time, it is observed during wire growth.
When the kink to $G=34\,G_0$ is nucleated, its amplitude is large enough that the overshoot ahead of the kink---which is now toward lower radii---is sufficient to trigger a new instability.
This forms an additional kink ahead of the existing one, as schematically shown in Fig.\ \ref{fig:kinkSchema}(d).
In that particular case, the dynamics is found to be dominated by the interkink force [Eq.\ (\ref{eq:Fkink})] only in the very early stages of evolution, while the external current quickly provides deviations from the $t^{1/2}$ behavior of the interkink distance.
Interestingly, the contribution of the outside current $J_0=J_l$ is attractive, so that the sign of the global force may change when the kinks are far apart.
As the system is overdamped, the interkink distance actually relaxes to an equilibrium distance, where the force vanishes (as long as no other soliton is nucleated.)

The general evolution of a system with multiple kinks and/or antikinks can be predicted by considering each pair of neighboring solitons separately.
A set of differential equations similar to Eq.\ (\ref{eq:kinkdist}) is thus obtained for each interkink distance and may be solved, together with an additional equation from global volume conservation, for each soliton position.
A general trend of the kink dynamics is to decrease their number and increase the length of cylindrical segments, as has been observed numerically\cite{burki03,burki04} in the equilibration of a random wire, leading to the universal equilibrium shape and the long-time evolution of an unstable wire.
In the latter case, the wire radius switches back and forth between two neighboring stable radii, so that Eq.\ (\ref{eq:Fkink}) simplifies to $F=\varepsilon(R_o)-\varepsilon(R_k)$, where $R_o$ and $R_k$ are, respectively, the radii outside and between the two kinks considered.
When $F<0$, the cylinder between the two kinks behaves exactly as a {\sl false vacuum}.

\section*{DISCUSSION AND CONCLUSIONS}\label{sec:concl}

The kink dynamics described in Sec.\ \ref{sec:singlekink} and \ref{sec:multkinks} is accessible to experiments and has, in fact, already been observed for gold\cite{oshima03,zandbergen05} and possibly for silver\cite{rodrigues02} wires in TEM experiments:
Movies of the real-time dynamics of gold wires \cite{TakayanagiMovies} show a stepwise thinning of a cylindrical wire from seven down to four atomic diameters through motion of kinks along the wire.
Solitons are seen to nucleate and move very rapidly---faster than the time resolution of the experiment---across a significant portion of the wire, in agreement with the fast initial propagation predicted during kink formation and overshoot expansion in the NFEM.
Some kinks are seen to stop along the way, as during the transition from six to four atomic diameters.
A second kink appears to join the first one before they move along. 
All this is in semiquantitative agreement with simulations presented here as the starting wire with a conductance $G=42G_0$ (Fig.\ \ref{fig:kinkpos}) has a diameter close to seven atomic diameters, while a wire with a conductance of $17G_0$ corresponds to five atomic diameters.
Note that conductance values quoted here are quantized values corresponding to ideal, ballistic transport.
Measured conductances\cite{oshima06} are likely to be lower due to backscattering of electrons\cite{burki99} from wire imperfections and disorder in the leads.

The simulation includes a sink of atoms, modeled as a constant boundary current.
This is meant to mimic a nonequilibrium situation, where the connection to the macroscopic leads allows for atoms to diffuse back and forth between the wire and the leads.
Experimentally, the wires are observed to thin down, presumably due to thermal fluctuations, electron irradiation from the electron microscope, and/or tension on the wire, thus the choice of a sink rather than a source of atoms.
A single sink of atoms is used in order to simplify the dynamics.
When two sinks of atoms are included, the dynamics is found to be dominated by the larger one:
Kinks nucleate at the position of the larger boundary current and propagate through most of the wire before the other sink nucleates the corresponding antikink.
The thinning dynamics with two unequal sinks of atoms is thus essentially equivalent to that of a single sink, except for the absorption at the wire end.
Finally, thermal fluctuations of the boundary current would provide fluctuations around the ideal dynamics discussed here but are not expected to alter it significantly.
The effect of thermal fluctuations of the whole wire structure has been studied using a stochastic dynamical model.\cite{burki05a}
The escape mechanism from long, metastable cylinders is found to be the nucleation of a kink at the wire end that propagates along the cylinder.
The effect of thermal fluctuations may thus be included to lowest order as a random fluctuation of the boundary current, which would make the thinning more ``jerky'' and thus even more similar to experiments.\cite{oshima03,TakayanagiMovies}

Material dependence is included in the model by adjusting the surface tension $\sigma_s$ in Eq.\ (\ref{eq:Omega_e}) to the appropriate bulk value.\cite{burki03,urban06,burki05}
The diffusive dynamics depends weakly on this parameter, except for a decrease of the critical length $L_c$ for nucleation of new solitons (see Sec.\ \ref{sec:multkinks}) with increasing surface tension.

The NFEM is a continuum model that exhibits some degree of discreteness, apparent through the formation of kinks connecting cylindrical parts of the wire.
Although the dynamics of the ionic background is fully classical in the NFEM, it is strongly influenced by the electron-shell potential, which is a result of the quantum confinement of the transverse motion of the electrons within the wire.
The latter provides a ``quantization'' of the continuous ionic structure on a scale given by the Fermi wavelength $\lambda_F$ and is responsible for the rich kink dynamics.

The NFEM assumes that the atomic structure of the wire adapts itself to the shape dictated by the electronic structure.
This hypothesis, as mentioned in the Introduction, is confirmed by the shell structure adopted by the atoms in thin nanowires.\cite{kondo00}
One of the main drawbacks of this assumption is that it neglects any possible back action of the atomic structure on the electronic motion.
However, these effects, though hard to evaluate, are expected to be minimal for the wires considered in this paper, which are thin enough that crystalline structure is not expected---and is indeed not observed \cite{kondo97,kondo00}---but thick enough for a continuum approximation to be reasonable.

The simplicity of the model makes simulations of large systems over long times tractable.
The initial wire for the simulation of Sec.\ \ref{sec:singlekink} corresponds to $4\cdot10^4$ atoms, over an evolution time of seconds.
This is far beyond what any simulation based on more ``realistic'' models, such as {\sl ab initio} methods\cite{tosatti01,dasilva04} or classical molecular dynamics,\cite{jagla01,koh06} can hope to achieve in the foreseeable future, as they are currently limited to at most a few nanoseconds.
Furthermore, classical molecular-dynamics simulations, which can treat relatively large systems, do not include electron-shell effects and are therefore unable to even stabilize a long nanocylinder.
The NFEM---with all its simplifying assumptions, and thanks to them---is to date the only model allowing realistic predictions of the long-time dynamics of metal nanowires.
As the latter is now  accessible experimentally with atomic resolution, these predictions are falsifiable and can be rigorously tested.

Finally, although the NFEM is, in principle, limited to simple free-electron-like metals, such as the alkali metals and to some extent noble metals, confinement of the electronic transverse motion is a very general feature of metals as soon as the transverse linear dimension of the system is of the order of the Fermi wavelength.
As such, its consequences, from the magic radii for the stability to the discreteness of the ionic dynamics, are expected to be quite general.

\section*{ACKNOWLEDGMENTS}

I am grateful to Charles Stafford for enlightening discussions on various aspects of this research. 
This work was supported by NSF Grants N0. 0312028 	
and No. 0351964.					

\bibliography{refs}

\begin{thebibliography}{38}
\bibitem[{Sur()}]{SurfAtoms}
\bibinfo{howpublished}{For example, more than $60\%$ of the atoms of a wire
  with a diameter of six atomic layers, corresponding to a conductance around
  $30G_0$, are at the wire's surface.}
\bibitem[{inc()}]{incompFluid}
\bibinfo{howpublished}{The incompressibility assumption is not critical.
  Alternative constraints have been considered (Ref.\ \onlinecite{urban04}) but
  do not significantly modify the results.}
\end{thebibliography}

\end{document}